\documentclass[11pt]{article}
\usepackage{wds11}

\usepackage[square]{natbib}

\usepackage{physics}
\usepackage{graphicx}       
\usepackage{xcolor}
\usepackage{hyperref}

\newcommand{\covd}[2]{ \frac{ {\rm D}{#1} }{ {\rm d}{#2} } }

\lefthead{SKOUPY AND LUKES-GERAKOPOULOS}
\righthead{GW TEMPLATES FROM EMRI}

\begin{document}

\title{Gravitational wave templates from Extreme Mass Ratio Inspirals}

\author{V. Skoupý}

\affil{Institute of Theoretical Physics, Faculty of Mathematics and Physics,
Charles University, CZ-180 00 Prague, Czech Republic}
\affil{Astronomical Institute of the Czech Academy of Sciences, Bo\v{c}n\'{i} II 1401/1a, CZ-141 00 Prague, Czech Republic}     

\author{G. Lukes-Gerakopoulos}

\affil{Astronomical Institute of the Czech Academy of Sciences, Bo\v{c}n\'{i} II 1401/1a, CZ-141 00 Prague, Czech Republic}

\begin{abstract}
An extreme mass ratio inspiral takes place when a compact stellar object is inspiraling into a supermassive black hole due to gravitational radiation reaction. Gravitational waves (GWs) from this system can be calculated using the Teukolsky equation (TE). In our case, we compute the asymptotic GW fluxes of a spinning body orbiting a Kerr black hole by solving numerically the TE both in time and frequency domain. Our ultimate goal is to produce GW templates for space-based detectors such as LISA.
\end{abstract}

\begin{article}

\section{Introduction}

The Laser Interferometer Space Antenna (LISA) is a future space based gravitational wave (GW) detector planned to launch in 2030s \citep{lisa}. LISA will consist of three spacecrafts forming an equilateral triangle with sides 2.5 million km long on heliocentric orbit. Changes in the distance between the spacecrafts will be monitored by Michelson-like interferometers with high precision. When a GW passes this constellation, the spacetime between the spacecrafts will be quasi-periodically stretched and contracted in the direction perpendicular to the propagation of the wave. Hence, LISA will be able to detect the wave, including its amplitude and phase, by measuring the changes in the distance between the three spacecrafts. This GW observatory will be sensitive in frequencies around $10^{-3}$ Hz.

One type of events LISA will be able to detect are extreme mass ratio inspirals (EMRIs). It is expected that the centre of galaxies host supermassive black holes with masses in the range of $10^6$--$10^9 M_\odot$. An EMRI takes place when a stellar mass black hole or a neutron star is inspiraling into a supermassive black hole while losing energy and angular momentum due to gravitational radiation reaction. Such a system is emitting GWs that should be detectable far away from the source in the mHz bandwidth. 

A GW signal from an EMRI can provide important information about the parameters of the system such as the masses of the objects, their spins etc. Actually EMRIs' detection will give us the opportunity to map the spacetime around a supermassive black hole to high accuracy. Since the parameter analysis of the detected GW signal depends on the accuracy of the waveform templates, it is important to model an EMRI with adequate precision and find theoretical waveforms from EMRI systems for various parameters.

The radiation reaction that acts on a moving particle can be split in two parts: a dissipative and a conservative one. In this work, we focus on the dissipative part, which can be calculated from the energy and angular momentum fluxes at the horizon of the black hole and at infinity. To obtain the leading term in the evolution of the GW phase, it is sufficient to consider only time averages of the dissipative part \citep{barack2019}. This approximation is called adiabatic. In this case, the particle is slowly shifted from one orbit to another on time scale much larger than the orbital period.

In this paper, we first review the properties of a spinning test particle moving in the Kerr spacetime. Then we summarize the Teukolsky formalism, which allows us to calculate the energy and angular momentum fluxes along with the waveforms. Using this formalism we calculate the energy fluxes from circular equatorial orbits of spinning particles around Schwarzschild and Kerr black holes. Subsequently, we use these fluxes to adiabatically evolve circular equatorial orbits and to find the effects of spin of the secondary object on the GW phase. Throughout the paper, we use geometrical units $G=c=1$,  where $G$ is the gravitational constant and $c$ is the speed of light, and the metric signature $({-}{+}{+}{+})$.

\section{Spinning test particles in the Kerr geometry}

The Kerr geometry, which describes a rotating black hole in vacuum, is represented in Boyer-Lindquist coordinates $(t, r, \theta, \varphi)$ by the metric  \citep{mtw}
\begin{equation}
    \dd s^2 = -\frac{\Delta}{\Sigma} \qty( \dd t - a \sin^2\theta \dd \varphi )^2 + \frac{\Sigma}{\Delta} \dd r^2 + \Sigma \dd \theta^2 + \frac{\sin^2\theta}{\Sigma} \qty( \qty(r^2+a^2) \dd \varphi - a \dd t )^2 
\end{equation}
where
\begin{equation}
\Delta = r^2-2Mr+a^2 \, , \qquad \Sigma = r^2+a^2 \cos^2\theta \, . 
\end{equation}
This metric depends on two parameters: the mass of the black hole $M$ and the Kerr parameter $a$. The internal angular momentum (spin) of a Kerr black hole is $a\,M$.  At the radius $r_+ = M + \sqrt{M^2 - a^2}$, where $\Delta = 0$, the outer event horizon is located. In this paper, we are dealing only with the region $r>r_+$. For $a=0$ the Kerr spacetime reduces to the Schwarzschild one.

A compact test object in general relativity can be characterized just by its multipoles \citep{dixon}. For example, a rotating black hole or a neutron star moving in a Kerr background can be modeled using a pole-dipole approximation, where only the mass and the spin of these compact objects are taken into account, reducing them to a spinning test particle. The pole-dipole approximation holds as long as the size of the test body is smaller than the scale of the background curvature. The stress-energy tensor of a spinning test particle reads \citep{ehlers}
\begin{equation}\label{eq:StrEn}
 T^{\mu\nu} = \frac{1}{\sqrt{-g}} \qty( \frac{ v^{(\mu} p^{\nu)} }{v^0} \delta^3 - \nabla_\alpha \qty( \frac{ S^{\alpha (\mu} v^{\nu)} }{v^0} \delta^3 ) )
\end{equation}
where $g$ is the determinant of the metric, $v^\mu=\displaystyle \dv{x^\mu}{\tau}$ is the four-velocity, $p^\mu$ is the four-momentum, $S^{\mu\nu}$ is the spin tensor,  $\delta^3 = \delta(r-r_p(t)) \delta(\theta-\theta_p(t)) \delta(\varphi-\varphi_p(t))$ where $r_t(t)$, $\theta_p(t)$ and $\varphi_p(t)$ are the coordinates of the particle depending at the coordinate time $t$. The conservation of the stress-energy tensor~\eqref{eq:StrEn} leads to the Mathisson-Papapetrou-Dixon (MPD) equations  \citep{mathisson,papapetrou,dixon}
\begin{align}
    \covd{p^\mu}{\tau} &= -\frac{1}{2} R^\mu{}_{\nu\kappa\lambda} v^\nu S^{\kappa\lambda} \, , \\
    \covd{S^{\mu\nu}}{\tau} &= p^\mu v^\nu - p^\nu v^\mu \, ,
\end{align}
where $\tau$ is the proper time and $R^\mu{}_{\nu\kappa\lambda}$ is the Riemann tensor.

The centre of mass for an extended body in general relativity is not uniquely defined. To fix the centre of mass for a spinning body, one has to specify the so called spin-supplementary condition (SSC). In this work we use the Tulczyjew-Dixon SSC \citep{dixon}, which reads
\begin{equation}
    S^{\mu\nu} p_\mu = 0 \, .
\end{equation}
and closes the MPD system. Actually, this SSC allows an explicit relation of the dependence of the four-velocity $v^\mu$ on the four-momentum $p^\mu$ \citep{ehlers77}. 

The magnitude of the spin is defined as $S^2 =  S^{\mu\nu} S_{\mu\nu}/2$, while the mass of the particle is $\mu^2=-p^\mu p_\mu$. Both of these quantities are conserved under Tulczyjew-Dixon SSC.  Instead of the measure of the spin $S$, its dimensionless counterpart $\sigma = S/(\mu M)$ is often used in EMRI studies. For this dimensionless spin holds $\sigma\simeq \mu/M\equiv q$, i.e. it is of the order of an EMRI mass ratio.

There are two Killing vectors in the Kerr geometry
\begin{equation}
\xi^\mu_{(t)} = \pdv{x^\mu}{t}\, , \qquad \xi^\mu_{(\varphi)} = \pdv{x^\mu}{\varphi} \, ,
\end{equation}
providing respectively two conserved quantities for the spinning particle
\begin{align}
 E &= - \xi^\mu_{(t)} p_\mu + \frac{1}{2} \xi_{\mu;\nu}^{(t)} S^{\mu\nu} \, , \\
 J_z &= \xi^\mu_{(\varphi)} p_\mu - \frac{1}{2} \xi_{\mu;\nu}^{(\varphi)} S^{\mu\nu} \, .
\end{align}
These quantities can be interpreted at infinity as the energy and the component of the total angular momentum parallel to the rotational axis of the central black hole ($z$-axis).

\section{Teukolsky equation}

The mass ratio $q$ of an EMRI lies between $10^{-7}$ and $10^{-4}$. Thanks to this, the GWs from such systems have relatively low amplitudes and perturbation theory can be employed. Using Newman-Penrose formalism it is possible to find equations governing  the perturbation of the Weyl tensor projected on some tetrad.

\cite{teukolskyI}  found the master equation (Teukolsky equation, TE) 
\begin{multline} \label{tkeq}
\qty( \frac{ \qty( r^2+a^2 )^2 }{\Delta} - a^2 \sin^2 \theta ) \pdv[2]{\psi}{t} + \frac{4Mar}{\Delta} \pdv{\psi}{t}{\varphi} + \qty( \frac{a^2}{\Delta} - \frac{1}{\sin^2 \theta} ) \pdv[2]{\psi}{\varphi}  \\
 - \Delta^{-s} \pdv{r}( \Delta^{s+1} \pdv{\psi}{r} ) - \frac{1}{\sin\theta} \pdv{\theta}( \sin\theta \pdv{\psi}{\theta} ) - 2s \qty( \frac{a\qty(r-M)}{\Delta} + \frac{i \cos\theta}{\sin^2 \theta} ) \pdv{\psi}{\varphi}  \\ 
 - 2s \qty( \frac{M\qty(r^2-a^2)}{\Delta} - r - i a \cos\theta ) \pdv{\psi}{t} + \qty( s^2 \cot^2 \theta - s ) \psi = 4\pi \Sigma T \, ,
\end{multline}
which governs scalar, neutrino, electromagnetic and gravitational perturbations of the Kerr spacetime. $s$ denotes the spin weight of the field and $\psi$ is a projection of the field quantity on a tetrad depended on $s$, while $T$ is the source term. For GWs at infinity it is useful to calculate the quantity $\Psi_4 = \rho^4 \psi$ for $s=-2$, where $\rho = -1/(r-ia\cos\theta)$. Then the source term consists of derivatives of the stress-energy tensor projected on the tetrad.

This equation is usually decomposed into azimuthal $m$-modes 
\begin{equation}
    \psi(t,r,\theta,\varphi) = \sum_{m=-\infty}^\infty \psi_m(t,r,\theta) e^{im\varphi} \, , \qquad \psi_m(t,r,\theta) = \frac{1}{2\pi} \int_{0}^{2\pi} \psi(t,r,\theta,\varphi) e^{-im\varphi} \dd \varphi \, .
\end{equation}
This replaces the derivatives $\dv*{}{\varphi}$ with $im$.

\subsection{Time domain}
It is possible to numerically solve the (2+1)dimensional TE in the time domain. We solve TE equation in the so called horizon-penetrating hyperboloidal (HH) coordinates $(\tau,\rho,\theta)$. Hypersurfaces of constant time in these coordinates are light-like at the horizon and null infinity and compactified in the radial direction, which allows us to deal with the boundary conditions. By defining the quantity $\phi = \partial_\tau \psi_m$ one can split the system into two first order in time and second order in space differential equations, which then can be solved by using the methods of lines \citep{harms2014}.

The energy flux $\dv*{E^\infty}{t}$ and the angular momentum fluxes at infinity can be calculated from the strain $h = h_+ - i h_\times$, where $+$ and $\times$ are the polarizations of the GW. The second derivative with respect to time of the strain is asymptotically proportional to the field quantity $\ddot{h} \propto \psi$ at infinity. The energy flux $\dv*{E^{\rm H}}{t}$ and the angular momentum fluxes can be calculated at the horizon as well.

\subsection{Frequency domain}
It is also possible to solve the TE (\ref{tkeq}) in the frequency domain by Fourier transformation in time, i.e.
\begin{equation}
    \psi(t,r,\theta,\varphi) = \sum_l \sum_{m=-l}^l \int_{-\infty}^\infty \dd\omega {}_sS^{a\omega}_{lm}(\theta) {}_sR_{lm\omega}(r) e^{-i\omega t + im\varphi}\, .
\end{equation}
By employing this transform separated ordinary differential equations can be derived for the spin weighted spheroidal harmonic function ${}_sS^{a\omega}_{lm}(\theta)$ and the radial function ${}_sR_{lm\omega}(r)$. This separation is an advantage of the frequency domain method, since it  highly reduces the computational cost. On the other hand, only GWs from bound multiperiodic orbits without dissipation can be calculated by a summation over discrete frequencies. Thus, different methods have to be employed to calculate the GW fluxes from inspiralling orbits.

The energy flux at infinity is given by \citep{drasco2006}
\begin{equation}\label{fluxinf}
 \dv{E^\infty}{t} = \sum_{l,m} \frac{\abs{Z^{\rm H}_{lm}}^2}{4\pi \omega_{m}^2}\, ,
\end{equation}
where the amplitudes $Z^{\rm H}_{lm}$ can be calculated as a convolution of the radial function $R^{\rm Up}_{lm\omega}$, which is regular at the horizon, and the source term derived from the stress-energy tensor (\ref{eq:StrEn}). For circular equatorial orbits of spinning particles with spin parallel to the $z$-axis at radius $r_0$ these amplitudes read
\begin{equation}
 Z^{\rm H}_{lm} = A_0 R^{\rm Up}_{lm\omega}(r_0) + A_1 \dv{R^{\rm Up}_{lm\omega}}{r} \qty(r_0) + A_2 \dv[2]{R^{\rm Up}_{lm\omega}}{r} \qty(r_0) + A_3 \dv[3]{R^{\rm Up}_{lm\omega}}{r} \qty(r_0)\,.
\end{equation}
We have derived independently the exact expressions of these coefficients $A_i$ and cross checked them with those of \cite{piovano2020}. This derivation is possible thanks to the fact that the orbital frequency $\Omega = \dv*{\varphi}{t}$ is constant for circular equatorial orbits, and, hence, each $m$-mode consists of only one frequency $\omega_m = m\Omega$. Similar expressions can be derived for the flux at the horizon and the angular momentum fluxes. We have checked that the frequency domain results agree for various orbits with the time domain approach results.

\begin{figure}[t]
\begin{center}
\begin{tabular}{cc}
\includegraphics[width=0.47\textwidth]{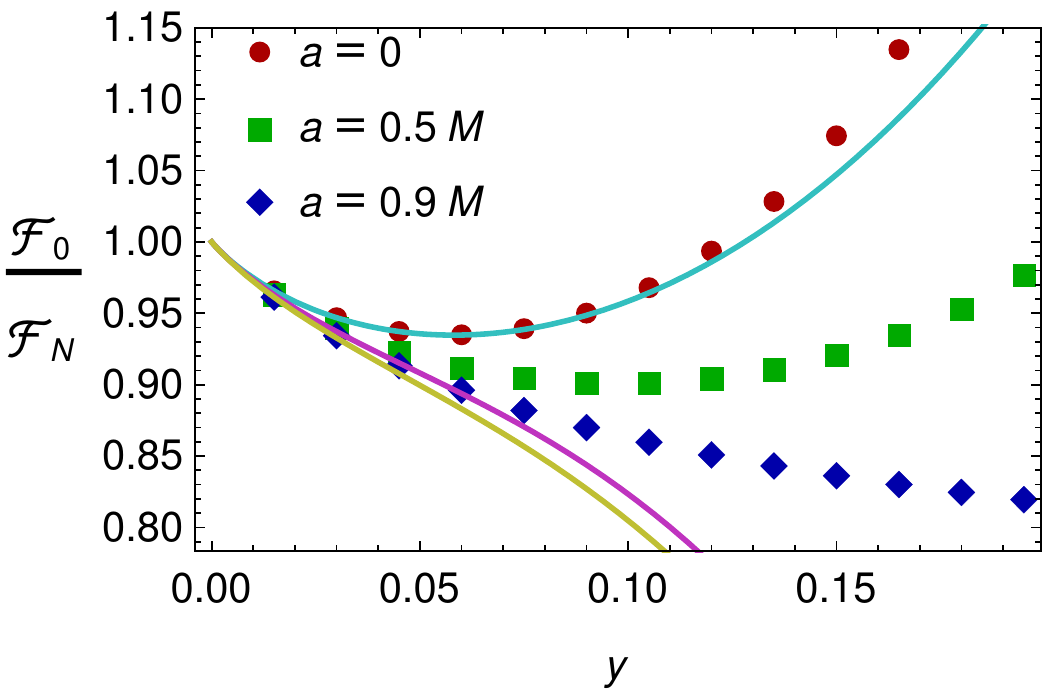} & \includegraphics[width=0.47\textwidth]{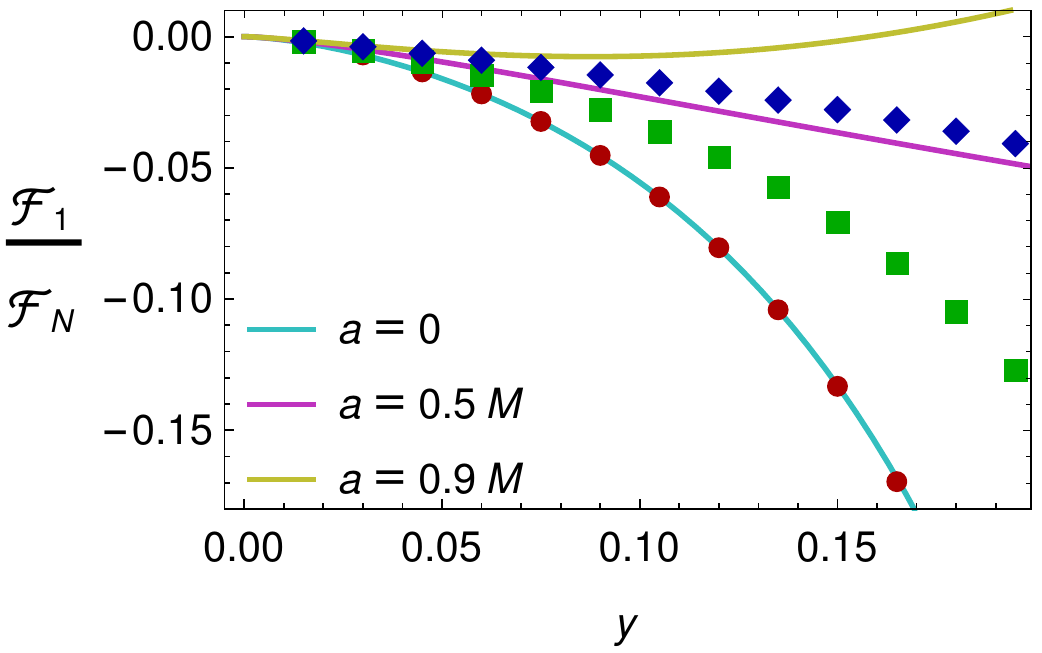} 
\end{tabular}
\end{center}
\caption{Dependence of total energy flux (\ref{flux}) normalized by the quadrupole formula $\mathcal{F}_N = 32 q^2 y^5/5$ \citep{tanaka1996} on the frequency parameter $y=(M\Omega)^{2/3}$. Dots indicate data calculated using the equations (\ref{fluxinf}) and (\ref{flux}), while solid lines show post-Newtonian (PN) results calculated using the BHPT \citep{Nagar:2019} (5.5PN for $a=0$ and 2.5PN for $a\neq 0$). The PN approximation fails for large $y$ especially for cases $a\neq 0$ because the PN order is lower.}\label{fig:fluxes}
\end{figure}

We have calculated the total energy fluxes of spinning particles moving on circular equatorial trajectories around a black hole with their spin parallel to the $z$-axis for several values of the frequency parameters $y\equiv (M\Omega)^{2/3}$. These calculations have taken place on a black hole background for $a=0$, $a=0.5M$ and $a=0.9M$, while the spin of the secondary $\sigma$ ranged between $-0.5$ and $0.5$ with step $0.1$. The dependence of the flux on the spin was fitted with a polynomial and the linear part was extracted to obtain
\begin{equation}\label{flux}
\mathcal{F} = \dv{E^\infty}{t} + \dv{E^{\rm H}}{t} = \mathcal{F}_0 + \sigma \mathcal{F}_1 + \order{\sigma^2}\, ,
\end{equation}   
where  $\mathcal{F}$ denotes the total energy flux, $\mathcal{F}_0$ is a constant term corresponding to the flux from a non-spinning particle and $\mathcal{F}_1$ is the term linear in spin. Both terms are plotted in Fig.~\ref{fig:fluxes}. The radial and angular functions and their derivatives were calculated using the Black Hole Perturbation Toolkit (BHPT) \citep{BHPToolkit}.

\section{Adiabatic inspiral}

A geodesic orbit in Kerr can be characterized by its constants of motion, i.e. the energy $E$, the $z$-component of the angular momentum $J_z$ and the Carter constant $Q$ \citep{schmidt2002}. For a spinning particle, the Carter constant is in general missing, it can be only retrieved when the MPD system is linearized in spin \citep{Witzany19}. Hence, when the particle is orbiting on the equatorial plane and its spin is parallel to the orbital angular momentum and $z$-axis, one can use energy $E$ and angular momentum $J_z$ to characterize the orbit. Actually, for circular equatorial orbits only one parameter such as energy, radius or orbital frequency is needed. 

Due to the conservation of the energy and the angular momentum, the change of these parameters must be opposite to the energy and angular momentum fluxes at the horizon and at infinity. The rate of change of the orbital frequency is
\begin{equation} \label{Omega}
    \dv{\Omega}{t} = -\frac{\mathcal{F}(\Omega)}{\dv{E}{\Omega}}\, .
\end{equation}
We have derived the dependence of energy $E$ on frequency parameter $y$ linear in spin $\sigma$ for Kerr black hole:
\begin{equation}
    E(y) = \frac{1-2 x y}{x^{3/2}\sqrt{2-x^3-3 x y}} - \sigma \frac{y^{3/2} \qty(x^3-1+xy)}{x^{9/2}\sqrt{2-x^3-3 x y}}\, ,
\end{equation}
where $x=\sqrt[3]{1-a\Omega}$. This result agrees with the equation (82) of \cite{harms2016} for $a=0$ and to the first order in spin with the equation (39a) of \cite{Hinderer:2013}.

\begin{figure}[htb]
\begin{center}
\begin{tabular}{c}
\includegraphics[scale=0.9]{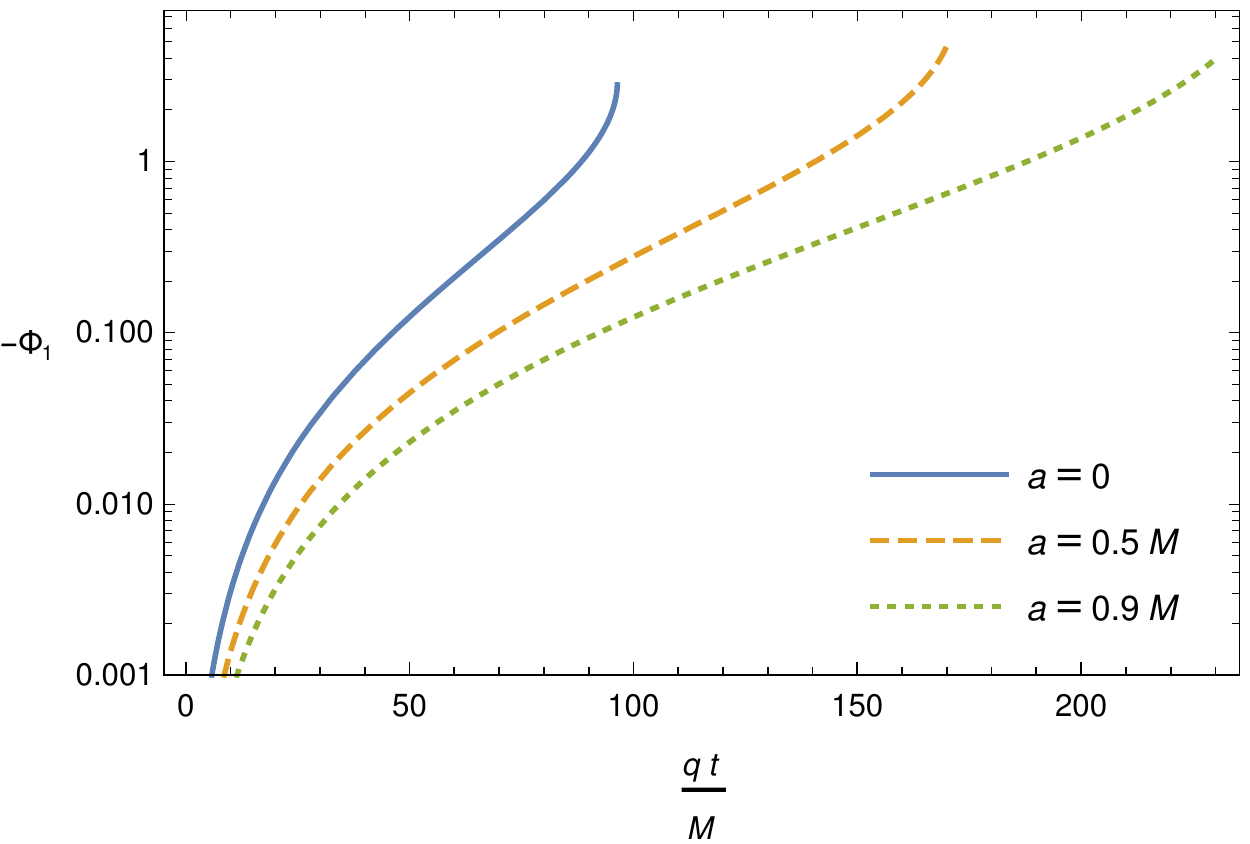}
\end{tabular}
\end{center}
\caption{Corrections to the GW phase caused by the spin of the secondary object. The initial frequency $\Omega$ is the same as the $\Omega$ for $r=10.1$ and given $a$. This plot is identical to the plot in Figure 3 of \cite{piovano2020}.} \label{deltaPhi}
\end{figure}

Because the spin $\sigma$ scales as the mass ratio $q$, the phase of the GW can be written as the following expansion \citep{piovano2020}
\begin{equation}
    \Phi(t) = \frac{1}{q} \Phi_0(t) + \frac{\sigma}{q} \Phi_1(t) + \order{\frac{\sigma^2}{q}} \, ,
\end{equation}
where the first term is of \emph{adiabatic} order and the second term is correction caused by the spin of the secondary object. The frequency of the $m$-modes is $m\Omega$ and the dominant mode is $m=2$. This implies that the GW phase is $\Phi(t) = 2\varphi(t)$. Suppose that the azimuth angle is $\phi(t) = \varphi_0(t) + \sigma \varphi_1(t) + \order{\sigma^2}$, then we can solve the system of equations $\dv*{\varphi}{t} = \Omega(t)$ and (\ref{Omega}) perturbatively to find that the correction to the phase is $\Phi_1(t) = 2q\varphi_1(t)$. To obtain an adiabatic inspiral the fluxes $\mathcal{F}_0$ and $\mathcal{F}_1$ from Fig.~\ref{fig:fluxes} were interpolated with a 3rd order Lagrange interpolation. The obtained results for $\Phi_1$ are shown in Fig.~\ref{deltaPhi} for different $a$; they are in agreement with those provided by \cite{piovano2020}, that have followed a different approach to obtain them.

\section{Conclusions}

Our main results are the following:
\begin{itemize}
    \item We have numerically calculated the energy fluxes from circular equatorial orbits of spinning particles with spin parallel to the $z$-axis. 
    \item We have used the above results to independently verify the results provided by \cite{piovano2020} by perturbatively solving the equations for the azimuthal angle $\varphi$ and the orbital frequency $\Omega$ to find correction to the GW phase caused by the spin of the secondary object. These results are shown in  Fig.~\ref{deltaPhi} and agree with those of \cite{piovano2020}. 
\end{itemize}

In a future work, these fluxes will be compared with fluxes computed using the time domain code (Teukode) for an inspiralling orbit to check whether the adiabatic approximation is justified.

\acknowledgments 
{Computational resources were supplied by the project "e-Infrastruktura CZ" (e-INFRA LM2018140) provided within the program Projects of Large Research, Development and Innovations Infrastructures and this work makes use of the Black Hole Perturbation Toolkit. The authors would also like to acknowledge networking support by the COST Action CA16104.}

\bibliographystyle{egs}
\bibliography{wds}       

\begin{thebibliography}{19}
\expandafter\ifx\csname natexlab\endcsname\relax\def\natexlab#1{#1}\fi
\expandafter\ifx\csname url\endcsname\relax
  \def\url#1{{\tt #1}}\fi
\expandafter\ifx\csname urlprefix\endcsname\relax\def\urlprefix{URL }\fi

\bibitem[{{Amaro-Seoane} et~al.(2017)}]{lisa}
{Amaro-Seoane}, P. et~al., {Laser Interferometer Space Antenna}, {\em arXiv
  e-prints\/}, p. arXiv:1702.00786, 2017.

\bibitem[{Barack and Pound(2019)}]{barack2019}
Barack, L. and Pound, A., {Self-force and radiation reaction in general
  relativity}, {\em Reports on Progress in Physics\/}, {\em 82\/}, 016\,904,
  \urlprefix\url{https://iopscience.iop.org/article/10.1088/1361-6633/aae552},
  2019.

\bibitem[{{BHPT contributors}(2020)}]{BHPToolkit}
{BHPT contributors}, {Black Hole Perturbation Toolkit},
  (\href{http://bhptoolkit.org/}{bhptoolkit.org}), 2020.

\bibitem[{Dixon(1964)}]{dixon}
Dixon, W.~G., {A covariant multipole formalism for extended test bodies in
  general relativity}, {\em Nuovo Cim\/}, {\em 34\/}, 317–339, 1964.

\bibitem[{Drasco and Hughes(2006)}]{drasco2006}
Drasco, S. and Hughes, S.~A., {Gravitational wave snapshots of generic extreme
  mass ratio inspirals}, {\em Physical Review D\/}, {\em 73\/}, 024\,027,
  \urlprefix\url{http://arxiv.org/abs/gr-qc/0509101}, 2006.

\bibitem[{Ehlers(1979)}]{ehlers}
Ehlers, J., {\em Isolated gravitating systems in general relativity:
  Proceedings of the International School of Physics "Enrico Fermi". Course
  67\/}, North-Holland Publ. Co, Amsterdam, 1. edn., 1979.

\bibitem[{{Ehlers} and {Rudolph}(1977)}]{ehlers77}
{Ehlers}, J. and {Rudolph}, E., {Dynamics of extended bodies in general
  relativity center-of-mass description and quasirigidity}, {\em General
  Relativity and Gravitation\/}, {\em 8\/}, 197--217, 1977.

\bibitem[{Harms et~al.(2014)Harms, Bernuzzi, Nagar, and
  Zenginoğlu}]{harms2014}
Harms, E., Bernuzzi, S., Nagar, A., and Zenginoğlu, A., {A new gravitational
  wave generation algorithm for particle perturbations of the Kerr spacetime},
  {\em Classical and Quantum Gravity\/}, {\em 31\/}, 245\,004,
  \urlprefix\url{https://iopscience.iop.org/article/10.1088/0264-9381/31/24/245004},
  2014.

\bibitem[{Harms et~al.(2016)Harms, Lukes-Gerakopoulos, Bernuzzi, and
  Nagar}]{harms2016}
Harms, E., Lukes-Gerakopoulos, G., Bernuzzi, S., and Nagar, A., {Spinning test
  body orbiting around a Schwarzschild black hole: Circular dynamics and
  gravitational-wave fluxes}, {\em Physical Review D\/}, {\em 94\/}, 104\,010,
  \urlprefix\url{https://link.aps.org/doi/10.1103/PhysRevD.94.104010}, 2016.

\bibitem[{Hinderer et~al.(2013)Hinderer, Buonanno, Mrou\'e, Hemberger,
  Lovelace, Pfeiffer, Kidder, Scheel, Szilagyi, Taylor, and
  Teukolsky}]{Hinderer:2013}
Hinderer, T., Buonanno, A., Mrou\'e, A.~H., Hemberger, D.~A., Lovelace, G.,
  Pfeiffer, H.~P., Kidder, L.~E., Scheel, M.~A., Szilagyi, B., Taylor, N.~W.,
  and Teukolsky, S.~A., Periastron advance in spinning black hole binaries:
  comparing effective-one-body and numerical relativity, {\em Phys. Rev. D\/},
  {\em 88\/}, 084\,005,
  \urlprefix\url{https://link.aps.org/doi/10.1103/PhysRevD.88.084005}, 2013.

\bibitem[{Mathisson(1937)}]{mathisson}
Mathisson, M., {Neue mechanik materieller systemes}, {\em Acta Phys. Polon.\/},
  {\em 6\/}, 1937.

\bibitem[{Misner et~al.(2017)Misner, Thorne, and Wheeler}]{mtw}
Misner, C., Thorne, K., and Wheeler, J., {\em Gravitation\/}, Princeton
  University Press, Princeton, 2017.

\bibitem[{Nagar et~al.(2019)Nagar, Messina, Kavanagh, Lukes-Gerakopoulos,
  Warburton, Bernuzzi, and Harms}]{Nagar:2019}
Nagar, A., Messina, F., Kavanagh, C., Lukes-Gerakopoulos, G., Warburton, N.,
  Bernuzzi, S., and Harms, E., {Factorization and resummation: A new paradigm
  to improve gravitational wave amplitudes. III. The spinning test-body terms},
  {\em Phys. Rev. D\/}, {\em 100\/}, 104\,056,
  \urlprefix\url{https://link.aps.org/doi/10.1103/PhysRevD.100.104056}, 2019.

\bibitem[{Papapetrou(1951)}]{papapetrou}
Papapetrou, A., {Spinning test-particles in general relativity}, {\em Proc. R.
  Soc. Lond.\/}, {\em 209\/}, 248–258, 1951.

\bibitem[{Piovano et~al.(2020)Piovano, Maselli, and Pani}]{piovano2020}
Piovano, G.~A., Maselli, A., and Pani, P., {Extreme mass ratio inspirals with
  spinning secondary: A detailed study of equatorial circular motion}, {\em
  Physical Review D\/}, {\em 102\/},
  \urlprefix\url{http://arxiv.org/abs/2004.02654}, 2020.

\bibitem[{Schmidt(2002)}]{schmidt2002}
Schmidt, W., Celestial mechanics in {K}err spacetime, {\em Classical and
  Quantum Gravity\/}, {\em 19\/}, 2743--2764, 2002.

\bibitem[{Tanaka et~al.(1996)Tanaka, Mino, Sasaki, and Shibata}]{tanaka1996}
Tanaka, T., Mino, Y., Sasaki, M., and Shibata, M., {Gravitational waves from a
  spinning particle in circular orbits around a rotating black hole}, {\em
  Progress of Theoretical Physics\/}, {\em 96\/}, 1087--1101, 1996.

\bibitem[{Teukolsky(1973)}]{teukolskyI}
Teukolsky, S.~A., Perturbations of a rotating black hole. {I}. {F}undamental
  equations for gravitational, electromagnetic, and neutrino-field
  perturbations, {\em The Astrophysical Journal\/}, {\em 185\/}, 635--647,
  1973.

\bibitem[{{Witzany}(2019)}]{Witzany19}
{Witzany}, V., {Hamilton-Jacobi equation for spinning particles near black
  holes}, {\em Physical Review D\/}, {\em 100\/}, 104\,030, 2019.

\end{thebibliography}

\end{article}
\end{document}